\documentclass[9pt]{book}

\usepackage[dvips]{graphicx,color}
\usepackage{makeidx,universe}

\makeauthorindex

\BookTitle{The proceedings of the Physics of Accreting Compact Binaries}

\CopyRight{\copyright 2010 by Universal Academy Press, Inc.}

\begin{document} 

\pagenumbering{arabic}

\chapter{%
Review Of The 2010 Eruption Of Recurrent Nova U Scorpii}

\author{\raggedright \baselineskip=10pt
{\bf Bradley E. Schaefer,
}\ \
{\small \it 
Department of Physics and Astronomy, Louisiana State University, Baton Rouge LA 70803, USA \\
}
}


\AuthorContents{Bradley E. Schaefer} 

\AuthorIndex{Schaefer}{B. E.} 

     \baselineskip=10pt
     \parindent=10pt

\section*{Abstract}

On 28 January 2010, the recurrent nova U Scorpii had its long predicted eruption; prior preparation allowed for this to become the all-time best observed nova event.  The coverage included daily and hourly spectra in the X-ray, ultraviolet, optical, and infrared, plus daily and hourly photometry in the X-ray, ultraviolet, U, B, V, y, R, I, J, H, K, and middle infrared, including roughly 35,000 V-band magnitudes (an average of better than once every three minutes) throughout the entire 67 days of the eruption.  This unprecedented coverage has allowed for the discovery of three new phenomema; the early fast optical flares (with no known explanation), ejecta velocities at 10,000 km/s (velocities that previously had only been seen in supernovae), and deep transient dips in optical and X-ray brightness lasting for hours (for which I point to X-ray dippers as having the same cause).

\section{U Scorpii}

U Scorpii (U Sco) is the third discovered recurrent nova (RN), with the discovery by H. Thomas of the 1906 and 1936 eruptions on the Harvard plates \cite{Thomas 1940} after the original 1863 eruption \cite{Pogson 1865}.  Ten known eruptions are now known, in 1863, 1906, 1917, 1936, 1945, 1969, 1979, 1987, 1999, and 2010 \cite{Schaefer 2010 ApJSupp}, with this review describing the 2010 eruption.  The eruptions rise very fast (within roughly six hours) to a peak of V=7.5, and then fade very fast, with the time to decline by three magnitudes being near 2.6 days \cite{Schaefer 2010 ApJSupp} or 3.6 days \cite{Schaefer et al. 2010 AJ}.  In either case, U Sco is the all-time fastest known nova event of any type.  The light curve shows a flat plateau phase from roughly 13-33 days after the peak, with this presumably caused by reprocessing of the light from the exposed supersoft source (SSS) due to residual nuclear burning close to the white dwarf (WD) as realized by Hachisu et al. \cite{Hachisu et al. 2000}.  In all previous eruptions, the light curve had not been followed after the plateau phase.  In quiescence, the nova was identified as a $B\approx19.2$ mag star \cite{Webbink 1978}.

I discovered that U Sco is an eclipsing binary with a deep eclipse and an orbital period of 1.23 days \cite{Schaefer 1990}.  This orbital period has been refined \cite{Schaefer and Ringwald 1995}, and now measured to high precision with 66 eclipses from 1989-2010 plus one eclipse from 1945 \cite{Schaefer 2010 eclipses}.  This eclipse nature of U Sco has allowed for critical and/or unique results:  (A) With this unusually long orbital period, the companion star must be a subgiant, and the accretion is driven by the evolutionary expansion of the star towards a red giant phase.  This expansion drives steady accretion at the high rate required to produce the short recurrence time scales seen in RNe.  (B) The eclipses are total, so at minimum we see only the unirradiated hemisphere of the companion.  The surface temperature of this hemisphere is measured by the colors at minimum and by the companion's spectral lines.  The size of the companion star is known from the orbital period, with little dependance on the mass of the companion star.  Along with the brightness at minimum, we can derive a blackbody distance of $12\pm2$ kpc \cite{Schaefer 2010 ApJSupp}.  This solves the old distance question, where distances from 3 to 96 kpc had been previously proposed.  Indeed, with the eclipses, U Sco now has one of the best known nova distances (only bettered for novae in clusters and external galaxies).  (C) The deep eclipses allow for precise minimum timings and high-accuracy orbital period measures.  With this, we can measure the steady orbital period change between eruptions due to mass transfer as well as the sudden orbital period change ($\Delta$P) across eruptions.  The $\Delta$P value is proportional to the mass ejected by the eruption ($M_{ej}$), which is a quantity that otherwise cannot be measured with useful accuracy.  This measure of $M_{ej}$ is critical for knowing whether the mass of the WD is growing or declining over the entire eruption cycle, and hence whether the system will become a Type Ia supernova.  RNe are one of the best candidate systems for the supernova progenitors, because they must have WDs near the Chandrasekhar mass and the accretion rate must be very high. As such, U Sco has a big part to play in the long-standing progenitor problem, with implications for the larger supernova cosmology programs.  (D) The deep eclipses allow for eclipse mapping of the region around the WD and in the disk.  This is particularly important during the eruption, when the eclipse mapping can be used to determine the fast-changing light sources in both the X-ray and optical.  The advance knowledge of the eclipses and their times allowed pre-eruption preparation for intense and detailed eclipse mapping, the first time such a program can be carried out.

\section{Prediction, Monitoring, and Discovery}

	The list of eruption years throughout the entire last century makes it obvious that U Sco erupts every $10\pm2$ years.  For this, there must have been missed eruptions around the years 1927 and 1957, despite extensive searches of archival plate collections.  We are actually lucky that only two eruptions have been missed in the last century.  U Sco is close to the ecliptic, with the Sun passing $3.6\deg$ away every 28 November; this can hide eruptions in a solar gap that can be pushed down to 62 days with extreme effort \cite{Schaefer et al. 2010 AJ}.  With U Sco being brighter than 14 mag for only 15 days during its eruption, eruptions are easily hidden in the solar gap.
	
	In 2005, I realized that the eruption dates of RNe can be predicted based on the quiescent B-band brightness level between eruptions \cite{Schaefer 2005}.  Any given RN will erupt only when a certain amount of mass has accreted, and this can be measured by monitoring the brightness level of the system.  For the RNe that do not have red giant companions, the blue system light is completely dominated by the accretion disk, so the accretion rate is proportional to a known power of the blue flux.  Integrations over many measures of the B-band flux will give a quantity proportional to the mass accreted since the last eruption.  Prior eruptions will give a measure of the mass required to trigger the eruption.  Empirically, U Sco is bright in quiescence (implying a relatively high accretion rate) when the inter-eruption interval is short, and faint when the interval is long.  With prior eruptions to set the threshold, we can predict the time of the next eruption as being when some known amount of B-band flux has accumulated since the last eruption.  On this physical basis, I predicted that U Sco would next erupt in 2009.3$\pm$1.0 \cite{Schaefer 2005}.  This is the first time that any nova event has been reliably predicted.
	
	With this strong prediction, in 2007 I started organizing a worldwide collaboration of observers over all wavelengths, for both photometry and spectroscopy.  Our group prepared detailed plans, arranged for many target-of-opportunity proposals, and recruited additional observers.  After extensive calculations, it was decided to not submit any proposals to HST or Spitzer (because neither a light echo nor the ejected shell had any chance of being detected) or to any of the biggest telescopes (because U Sco would be bright enough to be well observed with small- and moderate-sized telescopes and because the $\sim$5000 km/s expansion velocity gave no reason to need very high spectral resolution).  We tested our communications and developed a plan for rapidly alerting all observers.
	
	The speed of the outburst meant it was important to monitor U Sco frequently, so that the world could turn to the nova immediately and catch the peak.  To this end, we organized several systems of daily and hourly monitoring \cite{Schaefer et al. 2010 AJ}.  The backbone of this system was the many amateur astronomers of the American Association of Variable Star Observers (AAVSO).  I also had the four ROTSE telescopes (in Australia, Texas, Namibia, and Turkey) all taking hourly pictures whenever U Sco was above the horizon from any of the sites.  Further frequent monitoring was performed by Matthew Darnley and Michael Bode (Univ. Leicester) with the robotic 2.0-m Liverpool Telescope in the Canaries and 1.3-m SMARTS telescope on Cerro Tololo in Chile.  Special care was taken to close the solar gap as much as possible, and Shawn Dvorak even used the SOHO LASCO C3 instrument to look for eruptions in the ten-day interval centered on the solar conjunction. The AAVSO Headquarters performed the critical task of providing a 24/7/365 clearinghouse for reports, along with maintaining detailed instructions for alerting the world and invoking our many target-of-opportunity programs.  For the actual eruption, our organization worked perfectly.
	
	Barbara Harris (in New Smyrna Beach, Florida) was the AAVSO observer who first saw U Sco to be faint after it came out of its late-2009 conjunction with the Sun.  She kept looking each morning throughout January 2010 as U Sco was visible low in the dawn sky.  On 28 January, she was woken up by her pet dog, opened up her observatory (with a 16-inch Schmidt Cassegrain), and took two CCD images of the U Sco field.  On looking at her first image, she had the initial reaction that the field looked unfamiliar due to a bright star in the center, although she quickly realized that it was indeed U Sco in eruption.  Her discovery was not accidental, as she had been checking U Sco every morning for a long time with the goal of finding the eruption, and had already worked out lines of communication for just such a discovery.  Her first act was to immediately notify the AAVSO Headquarters, then she called me at home.  In the next five minutes, I found that the ROTSE telescope in Namibia had not taken a picture and no one else had reported the eruption.  In desperation to confirm the eruption (before pulling the trigger on getting the worldwide collaboration moving), I saw that the skies were clear over my home in Baton Rouge and U Sco was above the horizon, so I took my 6-inch telescope out into the front lawn.  With this, I made the confirmation that U Sco was in outburst.  After this, Shawn Dvorak (in Clermont, Florida) independently made his own discovery of the outburst.  The discovery of the U Sco eruption by both Harris and Dvorak demonstrated the importance of amateur astronomers, with this being a strong theme throughout the entire U Sco eruption.  Both Matthew Templeton (AAVSO) and Ashley Pagnotta (LSU) were woken up by Harris' alert, so we all started running down our long-prepared list for sending out circulars, notifying observers, and invoking prepared programs.
		
	All photometry for the prior pre-eruption interval is presented in \cite{Schaefer et al. 2010 AJ}.  No pre-eruption rise or dip was seen with limits of 0.2 mag in the few hours before the eruption.  The initial report of the discovery by Harris and Dvorak appeared in the IAU Circulars \cite{Schaefer et al. 2010a IAUCirc}.  The latest report of U Sco not in eruption is from JD 2455224.3438 by Yasunori Watanabe in Japan \cite{Schaefer et al. 2010 AJ}.  The best estimate is that the eruption started at JD 2455224.32$\pm$0.12, with a peak at V=7.5 at JD 2455224.69$\pm$0.07, and the discovery by Harris at JD 2455224.9385 (2010 January 28.4385 UT) \cite{Schaefer et al. 2010 AJ}.  For background and anecdotes on the discovery, see \cite{Simonsen and MacRobert 2010}.

\section{The All-Time Best Observed Nova Event}

We had three X-ray telescopes on satellites (Swift, RXTE, and INTEGRAL) in orbit looking at U Sco in the first day.  XMM, Chandra, and Suzaku all took long stares at U Sco on Days +9, +13, and +15 (with Suzaku, as reported by Dai Takei at this conference), on Day +18 (with Chandra, as reported by Marina Orio at this conference), and on Days +23 and +35 (with XMM, as reported by Jan-Uwe Ness at this conference), resulting in high time resolution X-ray light curves for a total of 4.1 days of exposure and high spectral resolution coverage from 0.2-10 keV on five epochs that span the eruption.  The Swift XRT made brief observations up to once every orbit throughout the entire eruption, providing an awesome record of the SSS's rise and fall and eclipse mapping of the X-ray source, and these observations can be binned together to give good low resolution spectra throughout the eruption.  INTEGRAL was watching U Sco for the first 13 days but did not see any hard X-ray emission, while the Swift BAT, RXTE, the Fermi GBM occultation program, and the MAXI detectors on the Space Station all looked throughout the entire eruption and saw no hard X-ray emission.  Astronomer's Telegrams for the U Sco X-ray observations are given by \cite{Manousakis ATel,Schlegel ATel1,Schlegel ATel2,Osborne ATel,Orio ATel,Schaefer ATel1}.

In the far ultraviolet, the Swift UVOT provided wonderful daily and hourly coverage with photometry and moderate-resolution spectroscopy (from 1700A into the visible band) throughout the entire eruption.  The XMM OM instrument (essentially identical to the Swift UVOT) provided fast time resolution UV photometry during the two three-quarter day stares on Days +23 and +35.

Classical photometry in a wide range of bands was carried out by many groups on a daily basis, which together gives full coverage from UBVRIJHK several times per day throughout all 67 days of the eruption:  BVRIJHK photometry was carried out by Ashley Pagnotta (LSU) with the 1.3-m SMARTS telescope on Cerro Tololo from the first day until the end of summer \cite{Pagnotta ATel}.  Gerald Handler (Univ. Vienna) had a 23-day observing run on the 0.5-m at SAAO in South Africa which fortuitously started at the time of the discovery, and he made nightly BVRI measures plus Stromgren b and y \cite{Pagnotta ATel}.  Ulisse Munari (INAF) used two small telescopes in Italy to get BVRI photometry from Days +1 to +40 \cite{Munari IBVS}.  Stromgren y-band photometry was made on many nights by both Seiichiro Kiyota (VSOLJ) and Hiroyuki Maehara (Kyoto Univ.), both of whom were present at the conference, as well as by James Clem (LSU) and Arlo Landolt (LSU) on many nights (going late into the eruption) with telescopes at Kitt Peak and Cerro Tololo.  JHK photometry nightly from Days +1 to +10 from Mount Abu in India are presented in \cite{Banerjee et al. 2010}.  In all, this full coverage allows for the creation of very broad spectral energy distributions on a daily basis throughout the eruption.

Fast photometry has been performed by many observers worldwide, now with a cumulative total of about 35,000 individual observations.  The bulk of this work has been done by well-equiped and skillful amateurs operating out of `backyard' observatories, taking long time series of CCD images.  Often these images are made without filters, and I find that a single constant offset for each observer makes their light curves fall exactly on top of simultaneous single points of classic V-band photometry.  The five PROMPT telescopes on Cerro Tololo (Aaron LaCluyze and Dan Reichart at Univ. North Carolina) each independently made simultaneous time series in U, B, V, R, and I throughout the eruption.  Hannah Worters, Ramotholo Sefanko, and Jaco Mentz (SAAO) made many time series in the V-band from Days +1 to +35.  The result of this overwhelming data set is that U Sco was followed about half of the entire 67 days of its eruption with fast photometry, with the gaps being primarily due to the Pacific and Indian Oceans, and overall with more than one magnitude every three minutes throughout the entire eruption

At this conference in Kyoto, many workers reported long series of spectra of U Sco during its eruption.  Masayuki Yamanaka (Hiroshima Univ.) reported on many optical spectra from telescopes around Japan \cite{Yamanaka et al. 2010}, while Kenji Tanabe and Kazuyoshi Imamura (Okayama Univ.) reported on their spectra of many novae including U Sco.  In addition, Kenzo Kinugasa (Gunma Astronomical Obs.) presented nine nights of optical spectra which show the complex evolution of the line profiles.  Due to the fortuitous location of Japan and the fast response of the Japanese observers, these spectral studies provide the earliest spectra of U Sco in eruption.  Michael Maxwell (Univ. Central Lancashire) presented a poster with infrared spectra from the 3.6-m NTT on La Silla, with these spectra already being published as part of a larger paper \cite{Banerjee et al. 2010} that includes infrared spectra from Mount Abu.  Elena Mason (ESO) presented a poster with her long series of optical spectra extending from early in the eruption until long after the end.  

Much optical and infrared spectroscopy has appeared in addition to those presented in Kyoto.  The single largest optical spectra data set is that of Fred Walter (SUNY Stony Brook) where he has nearly daily coverage over the entire eruption as taken with the CTIO 1.5-m telescope.  Stella Kafka (Carnegie) has a medium resolution spectrum taken with the Magellan 6.5-m telescope at Las Campanas from within a day of the discovery.  Howie Marion (Texas A\&M) and David Lynch (Aerospace Corp.) both have infrared spectra from the IRTF on Mauna Kea in Hawaii, on 29 January and 11 February respectively.  Michael Bode has a large number of optical spectra throughout the eruption with the 2.0-m Liverpool Telescope in the Canaries.  G. C. Anupama (IAPP) has 24 optical spectra from the first night all the way up until 10 April, all taken with the 2.0-m Himalayan Chandra Telescope in Kashmir \cite{Anupama 2010}.  These many data sets make for an optical spectrum series that has effectively daily coverage throughout the eruption and infrared spectrum series with good coverage up until the end of the plateau phase.

The WISE spacecraft scanned U Sco on two days starting on Day +28 with photometry in the 3.2 and 4.6 $\mu$ bands.  A. N. Ramaprakash (IUCAA) obtained multi-band
optical imaging polarimetry on Days +4 and +5 with the 2m IUCAA 
Girawali Telescope in India.  G. C. Anupama (IAPP) and S. P. S. Eyres (Univ. Central Lancashire) have used the Giant Metrewave Radio Telescope (near Pune, India) throughout the eruption and have made no detection.

The large numbers of observations, the coverage throughout the entire eruption, the fast cadence of spectra and photometry, and the coverage of all wavelength bands makes this eruption of U Sco into the all-time best observed nova event.

\section{Newly Discovered Phenomena} 
	With the unprecedented coverage and the first time that a nova has had any significant amount of fast photometry, several new phenomena were discovered for U Sco in 2010.  Here are three of these discoveries, plus one item that has been seen before but largely unrecognized.
\subsection{Early Flares} 
	
	Most novae generally have smooth monotonic declines after their peak \cite{Strope et al. 2010}.  U Sco is a P class light curve (i.e., with a plateau) which is taken to have a smooth light curve (other than for the eclipses after the plateau starts).  When the nova gets faint, we expect to see the usual flickering associated with the accretion disk.
	
	It came as a profound surprise when early in the eruption the fast photometry started to show large amplitude flares with a duration of around one hour.  These were first noted by \cite{Worters et al. 2010}, and many other observers recorded these flares.  The first was recorded on Day +6, and these continued until Day +13.  The amplitudes were from 0.1 mag to $>$0.5 mag.  The flare light curves appeared with different shapes, typically roughly triangular, with rise times of half an hour and durations of one hour.  Such flares have never been previously reported for a nova.  (However, at this conference, Hiroyuki Maehara reported that the U-Sco-like nova V2672 Oph also displayed similar flares on Days +6 and +8.)
	
	These early flares are {\it not} just a fast version of the jitters characteristic of the the J class novae (like DK Lac, HR Del, or V723 Cas).  The early flares in U Sco have a timescale that is two orders of magnitude faster than any jitters in J class novae, and there are no known flares with intermediate time scale to suggest any continuum of jitter durations \cite{Strope et al. 2010}.  Another good reason is that J class jitters always start before the peak and stop when the light curve has fallen by $\sim3$ mag from the peak \cite{Strope et al. 2010}, whereas for U Sco the flares start after the light curve has fallen by 5.5 magnitudes below peak.  Also, the J class novae are all very slow (all with $t_3>60$ days) \cite{Strope et al. 2010}, whereas U Sco is the all-time fastest known nova.
	
	These early flares are {\it not} the usual flickering, or anything else related to the inner binary or accretion disk.  The inner binary was shrouded in the SSS wind up until Day +15, as proven by the invisibility of the SSS emission from near the WD and by the lack of eclipses in the optical light curve.  So the early flares must be a phenomenon associated with the nova shell or the SSS wind.  
	
	With a risetime of half an hour, the flare emission region must be 30-light-minutes in size or smaller, and likely greatly smaller.  On Day +10, the expanding
shell (at velocities of 5000 km/s) is 4 light-hours in size.  So the
emitting region can only be a small fraction of the whole shell.  
With a flare amplitude of 0.5 mag, the flare emitting region is giving off
just a little less than the light from the entire shell.  The fast fall time implies that the `hot' mass must have a relatively small mass so that it can cool fast enough.  In all, the picture we 
have is a very small mass suddenly appearing, flaring, cooling fast, and 
fading away.  These fast flares have no precedent, prediction, or understanding from theory. 

	\subsection{Deep Aperiodic Dips in the Light Curve} 
	
	From Days +41 to +61, the out-of-eclipse light curve shows deep dips,
with typical depth 0.5-0.7 mag and typical duration 0.2 in phase.  These
dips occur at phases from 0.25-0.85, they show no apparent correlation
with orbital phase, and they do not recur from orbit to orbit.  The
dips are generally `V'-shaped.  These are big dips, highly
significant, and seen by multiple observers.  There is zero precedent for
anything like this from novae at any time (for the simple reason that no 
one has ever looked well enough in the past).

	I have a simple explanation for this new phenomenon.  The optical 
light comes primarily from a small region centered on the WD (as 
proved by the primary eclipses being deeper than 1.0 mag and being 
centered close to the time of conjunction).  The dips all start and stop 
near the regular out-of-eclipse level, so the dips can only be caused by 
eclipses of the central light source.  To produce eclipses at phases 
0.25-0.85, the eclipsing bodies can only be spread completely around the 
white dwarf.  Eclipse mapping over this time interval shows
that the optical light arises from a central-bright disk, so we know that 
the disk is present and spread around the WD, providing a source of 
occulters.  We also know the inclination of the U Sco orbit (either$\sim$80$^{\circ}$ \cite{Hachisu et al. 2000} or $82.7\pm2.9^{\circ}$ \cite{Thoroughgood et al. 2001}), so 
the line of sight to the central source is just skimming above the top of 
the accretion disk.  We also know that accretion disks can have raised 
rims caused by impacts of the accretion stream at various positions.  In the chaos as the 
accretion stream collides with itself and the disk, the disk will have 
transient raised rims and these will occult the bright inner 
regions, with the total optical light dipping.  This is a simple 
explanation using only well-known components.

	Indeed, there is a closely analogous case - the X-ray
dippers \cite{White and Swank,Frank et al. 1987}.  Many LMXRBs with inclination around 80$^{\circ}$ show X-ray dips which are transient with variable duration.  The X-ray dips are
certainly caused by raised rims in the accretion disk.  While there are
differences (a very small X-ray source near the neutron star versus a
larger optical source from the inner edge of the disk), the similarities 
are striking.  The X-ray dips provide a strong precedent and 
analogy for the optical dips.

	At this conference, Jan-Uwe Ness reports that the XMM light curve on Day +23 shows roughly 5 dips where the brightness falls to half-intensity, with the dips apparently restricted to phases within 0.25 of the optical eclipse.  At this time, the optical light curves do not show dips, while by Day +35 the XMM light curve shows a smooth X-ray eclipse with no dips.  The analogy with the classical X-ray dippers appears strong, so these X-ray dips are likely caused by raised rims of the accretion disk eclipsing the central light source.  (The lack of optical eclipses on Day +23 is simply due to the optical light coming from a large spherical source, as shown by the eclipse mapping, so any raised rim would occult only a small fraction of the whole optical brightness.  The optical dips start only when the optical light becomes dominated by the physically small inner regions of the disk around Day +41.)  Details for X-ray dip models have various uncertainties; for example, there is no X-ray dip centered on zero phase that could be the eclipse of the central X-ray source by the companion star, for which one possible explanation is that the rim of the accretion disk is quite high ($>$22$^{\circ}$) such that the secondary star only occults the outer edge of the accretion disk.

	\subsection{Expansion Velocities at 10,000 km/s} 
	
	Typical nova expansion velocities are less than 1000 km s$^{-1}$.  {\it All} RN have H$\alpha$ line profiles with FWHM$>$2000 km s$^{-1}$, while only 20\% of classical novae satisfy this condition \cite{Pagnotta et al. 2009}.  U Sco has previously been reported to have FWHM=8000 km s$^{-1}$ and FWZI=12,000 km s$^{-1}$ \cite{Iijima 2002}.  Now, \cite{Banerjee et al. 2010} have recognized that hydrogen and NI lines have wide and significant wings that extend out to 10,000 km s$^{-1}$, for a FWZI=20,000 km s$^{-1}$.  This paper examines and rejects the possibilities that the wings can be created by other lines and points to the same high velocity wings appearing in old spectra of the H$\alpha$.  These very high velocities only appear in the first several days of the eruption.  The 10,000 km s$^{-1}$ velocity is only the line-of-sight component, and it is possible that even higher expansion velocities are occurring (with U Sco having a bipolar outflow viewed from near the `equator', see \cite{Drake and Orlando}).  An expansion velocity of 10,000 km s$^{-1}$ is startling because it is so large for novae and is equal to the expansion velocity for supernovae.  No theoretical prediction or understanding of this unprecedented very high velocity is known.

	\subsection{Triple Peaked Line Profiles} 
	
	During the first five days of the U Sco eruption, the hydrogen line profiles are triple peaked, with a broad central component near zero velocity and two narrow components peaking around $\pm3000$ km s$^{-1}$.  The evolution of the three peaks is complicated \cite{Yamanaka et al. 2010}, with the two high velocity peaks declining with respect to the central peak.  From Days +16 to +23, the Balmer lines become triple again (with peaks near zero and $\pm1800$ km s$^{-1}$), with each peak being narrow and isolated.
	
	This triple-peaked structure has precedent in Type IIn supernova and in other novae.  The Type IIn SN1998S has similar triple-peaked profiles with the high velocity peaks attributed to an expanding torus, with the Earth being near the equatorial plane \cite{Gerardy et al. 2009}.  Fred Walter (SUNY Stony Brook) has found triple peaked profiles for YY Dor, Nova LMC 2009, V2672 Oph (see also \cite{Munari et al. 2010}), and KT Eri.  Triple-peaked profiles have also been reported for  V838 Her \cite{Vanlandingham et al. 1996}, V1187 Sco \cite{Lynch et al. 2006}, and V1494 Aql \cite{Kamath et al. 2005}, with all of these papers attributing the profiles to an expanding torus with the Earth being in the equatorial plane.

	Amongst these eight novae with triple peaked profiles, three of these are known RNe (U Sco, YY Dor, Nova LMC 2009), while three more are strongly suspected of being RN (V2672 Oph, KT Eri, and V838 Her).  So apparently, triple peaked profiles point to the recurrent novae, at least statistically.

        \cite{Drake and Orlando} calculate that the shell will look bipolar due to the accretion disk (like a pair of opposing mushroom clouds), with a low density region in the equatorial plane and a high density region around the rims of the mushroom clouds .  These two bright rims might be the expanding toruses that can provide the two high velocity peaks in the line profiles.  To see the redshifted high velocity peak, we have to see the backside of the torus.  During the first few days of the eruption while the ejected nova shell is optically thick, there must be some nearly-empty lines of sight so that we can see to the backside of the shell.  Given the known inclination of U Sco ($\sim$82$^{\circ}$), our line-of-sight to the back of the torus will pass just below the edge of the `upper' mushroom cloud, and so we'll have a relatively clear view of the receding part of the expanding torus.

\section{Detailed Evolution} 

{\bf Days 0 to +9.}  U Sco was discovered already fading, so the precise hour of maximum was missed by roughly 6 hours, with the gap due to oceans.  The decline was very fast, with estimated values of $t_2=1.7$ days and $t_3=3.6$ days \cite{Schaefer et al. 2010 AJ}.  This can be compared to estimates from prior eruptions that $t_2=1.2$ days and $t_3=2.6$ days \cite{Schaefer 2010 ApJSupp}, where it is unclear if these differences are caused by measurement errors or small-but-real differences between eruption.  U Sco is the fastest of all known nova events.  The speed implies that the mass of the white dwarf must be near the Chandrasekhar limit.  The decline is smooth up until Day +6.  During the first five days, the hydrogen line profiles are triple peaked.  During this interval, a faint X-ray flux is detected above 1 keV, with the likely cause being the nova shell ramming into some ambient gas.

{\bf Days +9 to +15.}  The overall light curve transitions from a fast decline to the start of the first plateau.  Starting on Day +6, U Sco shows large amplitude short-duration flares (see Section 4.1).  No eclipses are seen, so the shell, or more likely the wind being driven by the residual nuclear burning on the white dwarf, is still so optically thick that the central binary star is hidden in its mists.

{\bf Days +15 to +21.}  This interval is the start of the plateau, where the brightness levels off to V$\approx$14 mag.  Around Day +15 is when the SSS becomes bright and the first eclipses are visible.  Both these demonstrate that the nova shell and wind have decreased to the point where the region near the white dwarf is visible for the first time.  \cite{Hachisu et al. 2000} interpret the plateau as being caused by light from the fairly-steady SSS being reprocessed off material in the system.  During this interval, the brightness at phase 0.25 is $\sim$0.1 mag brighter than at phase 0.75, which is consistent with light from the SSS heating up the inner edge of the accretion stream.  Eclipse mapping is consistent with the optical light coming from a uniform sphere with radius 4.1 R$_{\odot}$.  The Balmer line profiles become isolated triple peaks.

{\bf Days +21 to +26.}  This interval is the middle of the first plateau.  The eclipses deepen to 0.8 mag.  The eclipse mapping is still consistent with the optical light coming from a spherical source with a radius a bit smaller than 4.0 R$_{\odot}$.  The asymmetry between phase 0.25 and 0.75 remains the same.  On Day +23, Jan-Uwe Ness reported at this conference that the XMM X-ray light curve shows deep dips (not seen in the optical or ultraviolet simultaneous light curves), with my interpretation being that these dips are caused by high edges of the circularization ring of the forming accretion disk.  The first classical flickering is seen on Day +24.5, and this proves that much of the accretion disk has reformed by this date.

{\bf Days +26 to +32.}  This interval is the last third of the first plateau, during which the eclipses deepen to 1.2 mag.  The eclipse mapping proves that most of the optical light is coming from an extended source that is elongated along the orbital plane, with a consistent configuration being an empty ring in the orbital plane.  This is what we would expect for a forming accretion disk where matter has not yet filled the interior.  The radius for this empty ring is 3.1 R$_{\odot}$.  Over this interval only, the secondary eclipse is visible in V-band light, implying that the secondary star's inward hemisphere has been greatly heated and is providing at least 20\% of the system's optical light.  The asymmetry between phase 0.25 and 0.75 continues.  The SSS continues at a fairly constant level, although getting a little brighter and hotter.

{\bf Days +32 to +41.}  The optical plateau ends with a sharp drop in brightness caused by the simultaneous turn off of the SSS.  The eclipses deepen to 1.4 mag.  Again, the eclipse mapping proves that the optical light is entirely coming from an elongated source in the plane of the orbit, with the distribution consistent with an empty ring of radius around 3.4 R$_{\odot}$.  On Day +35, the XMM light curve shows no dips and instead has a smooth symmetrical shape.

{\bf Days +41 to +54.}  The optical light curve flattens out at V$\approx$17 mag, forming a second plateau that has some ups and downs.  The eclipses become more shallow, with a depth of around 1.0 mag.  The shape of the eclipse light curves is similar to that during quiescence, with eclipse mapping showing a normal sized accretion disk that is centrally bright.  The filling in of the inner part of the accretion disk occurred in the few days before Day +41, likely because the cessation of the SSS and wind allowed for the inward migration of material in the circularization ring.  On Day +41, the first of the optical dips appears (see previous section), with my interpretation being that these are caused by temporary high edges of the accretion disk eclipsing the bright inner parts of the disk.

{\bf Days +54 to +67.}  This interval covers the final fading of the nova from the second plateau until the return to the quiescent level.  The last of the optical dips was seen on Day +61, so the outer edge of the accretion disk has settled down.  The asymmetry between phases 0.25 and 0.75 has vanished, and there is no secondary eclipse in the V-band.  The eclipse shape is largely the same as for the last time interval and as for quiescence.

\section{Looking Forward}

	Our community has a vast number of spectra in all wavelengths with awesome time coverage.  In following the mode of all previous previous nova spectral studies, individual authors are fully analyzing their own set of spectra, and this is good.  But we {\it also} have a wonderful opportunity to put {\it all} the spectra together into a grand global analysis, where the whole is much greater than the sum of the parts.   For example, a conclusion that the carbon abundance is
low would be weak if based on the X-ray spectra alone, yet would be strong
when confirmed with the UV, optical, and IR spectra.  Likely, the global
analysis will result in a better, more complex picture.

	To this end, Greg Schwarz (McMaster) and Robert Williams (STScI) have agreed to
perform this global analysis.   Already, they have obtained most of the world's spectra from X-ray to infrared.  The most important questions for the global analysis are the
abundances, the velocity structure, and the masses.  Is the accreted
matter really hydrogen deficient?  Do the ejecta contain any material
pointing to dredge-up?  If so, what does this imply about the composition
of the WD?  Can the many huge assumptions in the $M_{ej}$ calculation based 
on line fluxes be tested/confirmed/measured to get a measure of 
$M_{ej}$ that is better than the current $\sim$3-orders-of-magnitude accuracy?  
How do the complex and fast-changing triple peak structures in the line profiles change from line-to-line and how are these caused by the complex temperature/density structure of the bipolar ejecta?  What is the density as a function of velocity for the 
initial shell ejection and what are the velocity, temperature, and ejection 
rate as functions of time for the SSS wind?  Their global analysis should give unique and reliable results for many of the key questions on the physics 
of the nova ejecta.

	Another upcoming analysis of very high importance is direct measure of $M_{ej}$ based on the change of the orbital period across the eruption.  (The measures based on line fluxes are all uncertain by many orders of magnitude, while values based on theory have similar huge uncertainties even as one set of theoretical models cannot be used to `test' other models.)  Kepler's Law tells us that the orbital period will change as mass of the system changes due to its ejection of material.  With the conservation of angular momentum, the fraction change in the mass of the WD will be closely equal to the fractional change in the orbital period \cite{Schaefer and Patterson 1983}.  Since we closely know the mass of the WD (1.37 M$_{\odot}$), we can get a reliable measure of the ejected mass (independent of distance, extinction, filling fraction, temperatures, etc.) by a purely dynamical method.  With the reliable $M_{ej}$ plus the well-measured mass accreted between eruptions, we can know whether the WD is gaining or losing mass.  With this, we then answer the primary question as for whether U Sco (and the other long-period RNe) will `soon' have the WD explode as a Type Ia supernova.
		
		U Sco is a key system for understanding RNe and whether they will become Type Ia supernovae.  With eruptions happening every $10\pm2$ years, we can expect another very well observed nova event around the year 2020.



\end{document}